\documentclass[onecolumn,aps,prd,preprintnumbers,superscriptaddress,nofootinbib,amsmath,amssymb,floats,floatfix,showkeys,notitlepage,showpacs]{revtex4-1}

\usepackage{orcidlink}
\usepackage{comment}
\usepackage{lipsum}
\usepackage{graphicx}
\usepackage{subfigure}
\usepackage{palatino}
\usepackage{sans}
\usepackage{hyperref}
\hypersetup{colorlinks=true,linkcolor=blue,urlcolor=blue,citecolor=blue}
\usepackage[toc,page]{appendix}
\usepackage[normalem]{ulem}
\usepackage{adjustbox}
\usepackage{latexsym}
\usepackage{amsmath}
\usepackage{amssymb}
\usepackage{amsfonts}
\usepackage{dcolumn}
\usepackage{bm}
\usepackage{tikz}
\usepackage{bigints}
\usepackage{array,tabularx,multirow,booktabs}
\usepackage[tracking=true]{microtype}
\usepackage{multirow}
\SetTracking{}{500}
\SetTracking{encoding={*}, shape=sc}{40}
\UseRawInputEncoding 
\allowdisplaybreaks

\begin{document} \sloppy
\title{Traces of quantum fuzziness on the black hole shadow and particle deflection in the multi-fractional theory of gravity}

\author{Reggie C. Pantig \orcidlink{0000-0002-3101-8591}}
\email{rcpantig@mapua.edu.ph}
\affiliation{Physics Department, Map\'ua University, 658 Muralla St., Intramuros, Manila 1002, Philippines}

\begin{abstract}
In this paper, we investigate the properties of black holes within the framework of multi-fractional theories of gravity, focusing on the effects of q-derivatives and weighted derivatives. These modifications, which introduce scale-dependent spacetime geometries, alter black hole solutions in intriguing ways. Within these frameworks, we analyze two key observable phenomena - black hole shadows and particle deflection angle in the weak field limit - using both analytical techniques and observational data from the Event Horizon Telescope (EHT) for M87* and Sgr A*. The study from $q$-derivative formalism reveals that the multi-scale length $\ell_*$ influences the size of the black hole shadow in two ways, and modifies the weak deflection angle. Constraints on $\ell_*$ are derived from the EHT observations, showing significant deviations from standard Schwarzschild black hole predictions, which range from $10^{9}$ to $10^{10}$ orders of magnitude. Additionally, the weak deflection angle is computed using the non-asymptotic generalization of the Gauss-Bonnet theorem, revealing the effects of finite-distance and multi-scale parameters. Using the Sun for Solar System test, the constraints for $\ell_*$ range from $10^{8}$ to  $10^{9}$ orders of magnitude. Results from the weighted derivative formalism generates a dS/AdS-like behavior, where smaller deviations are found in the strong field regime than in the weak field regime. The results suggest that while these effects are subtle, they provide a potential observational signature of quantum gravity effects. The findings presented here contribute to the broader effort of testing alternative theories of gravity through black hole observations, offering a new perspective on the quantum structure of spacetime at cosmological and astrophysical scales.
\end{abstract}


\maketitle


\section{Introduction}\label{intr}
The exploration of black holes has long captivated the field of theoretical physics, not only for their dramatic properties within general relativity but also for the possibility they present in testing fundamental theories beyond Einstein's framework. In particular, black holes in higher dimensions provide a compelling arena for investigating how gravity behaves when the standard four-dimensional picture is expanded \cite{Myers:1986un, Emparan:2008eg}. Higher-dimensional spacetimes, as explored in the context of string theory \cite{Witten:1995ex, Maldacena:1997re,Randall:1999vf} and braneworld models, introduce novel features into black hole physics, such as modified horizons, thermodynamics \cite{Strominger:1996sh}, and new forms of gravitational radiation, which give some clue about the subtle and non-trivial structure of spacetime and open a pathway to probe alternative gravitational theories \cite{Chodos:1980df,Xu:1988ju,Shen:1989uj,Iyer:1989nd,Cadeau:2000tj,Myung:2013cna,Feng:2015jlj,Do:2019jpg,Cai:2020igv,Paul:2023gzj}.

Black holes within the framework of multi-fractional theories of gravity \cite{Calcagni:2016azd}, specifically focusing on two models: those with $q$-derivatives and weighted derivatives has been explored \cite{Calcagni:2017ymp}. In these models, spacetime dimensionality changes with the scale being probed, influenced by a fundamental length, $\ell_*$, and such modifications to the fabric of spacetime yield intriguing deformations to black hole solutions. In particular, Schwarzschild black holes within these multi-fractional contexts can exhibit altered event horizons and singularities. For instance, in the q-derivatives scenario, an additional ring singularity emerges, and the black hole's Hawking temperature can be higher than in general relativity. In the context of higher-dimensional black holes, its importance lies in its demonstration of how the multi-fractional geometry can modify black hole characteristics even at macroscopic scales, such as shifts in the event horizon, thermodynamic properties, and the presence of novel singularities. These effects, although subtle, provide an exciting opportunity to test phenomenological aspects of quantum gravity \cite{AlvesBatista:2023wqm}. The exploration of black holes within multi-fractional models, then, can contribute to a deeper understanding of how deviations from standard four-dimensional general relativity could manifest in observable astrophysical phenomena, providing insights into quantum gravity frameworks.

One of the most observable aspects of black holes lies in their shadows—regions formed by the gravitational bending of light near the event horizon, which is the photonsphere \cite{Claudel:2000yi,Virbhadra:2002ju}. The study of black hole shadows has evolved from theoretical predictions \cite{Synge:1966okc,Luminet:1979nyg,Falcke:1999pj} to direct observational tests, such as those conducted by the Event Horizon Telescope, which captured the shadow of the supermassive black hole M87* and Sgr. A* \cite{EventHorizonTelescope:2019dse,EventHorizonTelescope:2019ths, EventHorizonTelescope:2022xqj,EventHorizonTelescope:2022wkp,EventHorizonTelescope:2022wok}. Shadows are not only dependent on the mass and spin of the black hole but can also reveal signatures of the underlying gravitational theory \cite{Kumar:2018ple,Wei:2019pjf}. Due to such importance, it has been the favorite of many authors in the literature to analyze the black hole shadow's radius. In the context of multi-dimensional black holes, the reader is invited to these published references \cite{Papnoi:2014aaa, Singh:2017vfr,Nozari:2024jiz,Lemos:2024wwi,Roy:2023ine,Nozari:2023flq,Singh:2023ops,Mandal:2022oma,Uniyal:2022xnq,Banerjee:2022jog,Younesizadeh:2022xzy,Tang:2022hsu,Belhaj:2022cdh,Guo:2020nci,Belhaj:2020rdb,Ahmed:2020jic}, and some incomplete works \cite{Lobos:2024fzj,Novo:2024wyn}.

Another important black hole phenomenon that one can consider is the deflection angle, which has two regimes: weak and strong fields. The former measures the bending of light by black holes at larger impact parameters and is particularly sensitive to deviations from general relativity. That is, observing relativistic image lensing formations can provide limitations on the compactness of massive, dark entities, which is independent of their mass and distance \cite{Virbhadra:1998dy,Virbhadra:1999nm,Virbhadra:2007kw,Virbhadra:2008ws,Adler:2022qtb,Virbhadra:2022ybp,Virbhadra:2022iiy}. Using tools like the Gauss-Bonnet theorem \cite{Gibbons:2008rj}, which links the geometry of the spacetime to its topology, one can compute deflection angles that may differ in higher-dimensional or modified gravity settings. It is only recently, that the GBT was extended to include the finite distance of the source (of photons) and the receiver \cite{Ishihara:2016vdc}. Furthermore, its generalization where the integration domains were modified to include the photonsphere, was artfully considered \cite{Li:2020wvn}. It has been used to explore the deflection angle which is also generalized to include massive time-like particles \cite{Li:2019vhp,Li:2019qyb,He:2020eah,Li:2020ozr,Li:2021qei,Liu:2022hbp}.

It is the aim of this paper to probe whether the multi-fractional theory leaves measurable traces on these observable phenomena. Specifically, we investigate how a parameter emerging from the multi-fractional framework affects the black hole shadow and the weak deflection angle when viewed in the context of higher-dimensional black holes. Such an approach not only serves as a test of the multi-fractional theory itself but also enhances our understanding of the broader implications of non-standard dimensionality in gravitational systems. By examining these effects through established techniques like the non-asymptotic generalization of the Gauss-Bonnet theorem, we aim to bridge the gap between theoretical predictions and potential observational signatures, pushing forward our understanding of the deep structure of spacetime. At this time of writing, there are no studies in the literature exploring the multi-fractional nature of a black hole spacetime.

The paper is organized as follows: In Sect. \ref{sec1}, we briefly review the application of multi-fractional theory to black hole solutions. Then, in Sect. \ref{sec2}, we study the shadow and find constraints to the multi-scale length $\ell_*$ using the EHT results for M87* and Sgr. A*. In Sect. \ref{sec3}, we test the weak field regime for the parameter $\ell_*$. Finally, in Sect. \ref{conc}, we state final remarks and possible research prospects. Unless not specified differently, we work with the metric signature $(-,+,+,+)$ and geometrized units by setting $G = c = 1$.

\section{Brief review of a black hole in multi-fractional theory}\label{sec1}
In this section, we first review the black hole solution in multi-fractional theory with q-derivatives. In their seminal work \cite{Calcagni:2017ymp}, the standard General Relativity (GR) equations were modified, replacing ordinary derivatives with q-derivatives, which reflect a scale-dependent structure of spacetime. The first modification appears in the Riemann tensor, where the usual derivative \( \partial_\mu \) is replaced by the q-derivative:
\begin{equation}
    ^q R^\rho_{\mu \sigma \nu} = \frac{1}{v_\sigma} \partial_\sigma {}^q\Gamma^\rho_{\mu \nu} - \frac{1}{v_\nu} \partial_\nu {}^q\Gamma^\rho_{\mu \sigma} + {}^q\Gamma^\tau_{\mu \nu} {}^q\Gamma^\rho_{\tau \sigma} - {}^q\Gamma^\tau_{\mu \sigma} {}^q\Gamma^\rho_{\nu \tau},
\end{equation}
where \(v_\mu = \partial_\mu q_\mu(x_\mu)\) is a function that accounts for the measure dependence. The Christoffel symbols are also then modified as
\begin{equation}
    {}^q\Gamma^\rho_{\mu \nu} = \frac{1}{2} g^{\rho \sigma} \left( \frac{1}{v_\mu} \partial_\mu g_{\nu \sigma} + \frac{1}{v_\nu} \partial_\nu g_{\mu \sigma} - \frac{1}{v_\sigma} \partial_\sigma g_{\mu \nu} \right),
\end{equation}
and directly impact the geometry of spacetime in the multi-fractional framework. The Einstein-Hilbert action is generalized by introducing the measure \(v(x)\), which is a product of the q-derivative terms:
\begin{equation}
    {}^q S = \frac{1}{2 \kappa^2} \int d^D x \, v(x) \sqrt{-g} \left( {}^q R - 2 \Lambda \right) + S_m,
\end{equation}
where \( v(x) = \prod_\mu v_\mu(x_\mu) \) and \( {}^q R \) is the $q$-version of the Ricci scalar. This modification introduces a background scale dependence into the action.

It is then found that the solutions to Einstein’s equations look the same as in GR when written in terms of q-coordinates \( q_\mu \) \cite{Calcagni:2017ymp}. However, non-linear modifications appear when re-expressed in physical coordinates \( x_\mu \). It is important to emphasize that this is not simply a coordinate transformation (which one may confuse) but reflects the intrinsic multi-scale geometry of the theory. The Schwarzschild solution, then, in the multi-fractional theory, is
\begin{equation} \label{q_met}
    {}^q ds^2 = -\left(1 - \frac{2M}{q} \right) dt^2 + \left(1 - \frac{2M}{q} \right)^{-1} dq^2 + q^2(d\theta^2 + \sin^2 \theta d\phi^2),
\end{equation}
where \( q \) is the modified radial coordinate and where the non-trivial aspect of multi-fractal theory is hidden:
\begin{equation} \label{q}
    q = r \pm \frac{\ell_*}{\alpha}\left( \frac{r}{\ell_*} \right)^\alpha.
\end{equation}
We can see that inherently, $q$ is a function of the radial coordinate $r$. Furthermore, the length scale parameter $\ell_*$ marks the transition between UV and IR behavior of the geometry. When $\alpha = 1/2$, $ \ell_*  = l_{\rm Pl}^2/s$, where $s$ is called the observation scale length \cite{Calcagni:2016azd,Amelino-Camelia:2017pdr}. Indeed,  the parameter $\ell_*$ is the boundary at which the difference in physical laws occurs. It is further explained in \cite{Calcagni:2017ymp} that $q$ represents the stochastic fluctuation as can be gleaned from the upper and lower signs of the equation. The former simply represents the deformation in the radius, while the latter is about the quantum stochastic feature, where the radius suffers fuzziness due to multi-fractional effects of spacetime.

A different approach to multi-fractional theory is called weighted derivatives. It was shown that in the Einstein frame, the gravitational action in the theory with weighted derivatives
\begin{equation}
    S_g = \frac{1}{2 \kappa^2} \int d^4 x \sqrt{-\bar{g}} \left( \bar{R} - \omega \partial_\mu \Phi \bar{\partial}^\mu \Phi - e^{-\Phi} U \right),
\end{equation}
where \( \Phi(x) = \ln v(x) \) and \( U \) is a potential related to the measure weight \( v(x) \). Assuming an Ansatz,
\begin{equation}
    ds^2 = -\gamma_1(r) dt^2 + \gamma_2(r) dr^2 + r^2 (d\theta^2 + \sin^2 \theta d\phi^2),
\end{equation}
the field equations can be solved:
\begin{align}
    0 &= (\gamma_1 \gamma_2)', \nonumber \\
    0 &= \gamma_1'' - \gamma_1' \left(\frac{\gamma_2'}{2\gamma_2} + \frac{1}{r} - \frac{\gamma_1'}{2\gamma_1} \right) - \frac{\gamma_1}{r^2} \left( r\frac{\gamma_2'}{\gamma_2} - 2\gamma_2 + 2 \right),
\end{align}
where the solution is obtained as
\begin{equation} \label{gamma}
    \gamma_1(r) = 1 - \frac{2M}{r} - \frac{\chi r^2}{6}, \qquad \gamma_2(r) = \gamma_1(r)^{-1}.
\end{equation}
Here, $\chi$ can be treated as positive (dS-like), or negative (AdS-like). It is the simplest black hole solution where $\Lambda = (1/2)\chi$, an increase by a factor of $1/2$ due to the multi-scaling nature of the geometry - an alternative view for the cosmological constant as explained in \cite{Calcagni:2017ymp}.

\section{Analysis of shadow with effects from multi-fractional theory} \label{sec2}
For the calculations to be clear enough, we can still rewrite Eq. \eqref{q_met} as 
\begin{equation} \label{q_met2}
    {}^q ds^2 = A(q) dt^2 + B(q) dq^2 + C(q) d\phi^2,
\end{equation}
where we specialize at $\theta = \pi/2$, converting it to a $(2+1)$ metric. To derive the black hole shadow radius, we only need the expression for the photonsphere radius, and the critical impact parameter. The formalism we follow is well-known and used widely in the literature (see Refs. \cite{Claudel:2000yi,Virbhadra:2002ju,Perlick:2021aok}). Using Eq. \eqref{q_met2}, the photonsphere radius can be solved simply as
\begin{equation} 
    \frac{A(q)}{dq}q^2 - 2A(q)q = (2q - 6M) = 0,
\end{equation}
resulting to
\begin{equation}
    q(r_{\rm ph}) = 3M.
\end{equation}
Indeed, as one noticed, there is no need to implement some coordinate transformation in Eq. \eqref{q_met2} because the expression directly came from the Einstein-Hilbert action in the q-derivative formalism \cite{Calcagni:2017ymp}. As we set $\alpha = 1/2$ and choose the upper sign in Eq. \eqref{q}, we find two solutions for the photonsphere:
\begin{align} \label{rph}
    r_{\rm ph} = 2 \ell_* \mp 2 \sqrt{\ell_*(3 M +\ell_*)}+3 M.
\end{align}
We remark that if we have chosen the fuzziness feature of spacetime in Eq. \eqref{q}, we still obtain the same expression above. As we follow the reconciliation of the so-called problem of \textit{presentation} in \cite{Calcagni:2017ymp}, especially in their treatment of the black horizon, the upper sign in Eq. \eqref{rph} then represents the initial point presentation of the photonsphere, while the lower sign represents that of the final point. Remarkably enough, both of these cases are subjected to certain approximations that we can study. These are: (a) $\ell_*/M \to 0$, and (b) $\ell_*/M \to \infty$. It would mean that for the former, $M \gg \ell_*$ and for the latter, $\ell_* \gg M$. Finally, it is worthwhile examination of Eq. \eqref{rph} shows that it is reducible to the standard Schwarzschild case without the influence of the multi-scale length $(\ell_*=0)$.

Choosing the upper sign in Eq. \eqref{rph}, we find the following approximations for $\ell_*/M \to 0$ and $\ell_*/M \to \infty$ as
\begin{align} \label{rph-}
    r_{\rm ph}^{-} &\sim 3M - 2\sqrt{3M\ell_*} + 2\ell_* -  \mathcal{O}(\ell_*^{3/2}), \nonumber \\
    r_{\rm ph}^{-} &\sim \frac{9M^2}{4\ell_*} - \frac{27M^3}{8\ell_*^2} + \mathcal{O}(\ell_*^{-3}),
\end{align}
respectively. Choosing the lower sign, the two above approximations give
\begin{align} \label{rph+}
    r_{\rm ph}^{+} &\sim 3M + 2\sqrt{3M\ell_*} + 2\ell_* + \mathcal{O}(\ell_*), \nonumber \\
    r_{\rm ph}^{+} &\sim 4\ell_* + 6M - \frac{9M^2}{4\ell} + \mathcal{O}(\ell_*^{-2}),
\end{align}

With these simplified expressions for the photonsphere radii, we can now derive the impact parameter as
\begin{equation} \label{imp}
    b(r)^2 = \frac{C(q)}{A(q)} = \frac{\left(2 \sqrt{r\ell}+r \right)^{3}}{2 \sqrt{r\ell}\mp 2 M \pm r},
\end{equation}
and the critical impact parameter is simply using Eqs. \eqref{rph-}, and \eqref{rph+} on Eq. \eqref{imp}. That is, $b_{\rm crit} = b(r = r_{\rm ph})$. It turns out that if we use Eq. \eqref{rph+}, we need to use the upper in Eq. \eqref{q}. As for the lower sign, Eq. \eqref{rph-} is used. We found that it produces the same critical impact parameters. For the case of $\ell_*/M \to 0$ and $\ell_*/M \to \infty$, we find
\begin{align} \label{bcrit}
    b_{\rm crit}^2 &\sim 27M^2 + \frac{3 \ell_*^2}{M} + \mathcal{O}(\ell_*^{7/2}) \nonumber \\
    b_{\rm crit}^2 &\sim 27M^2 + \frac{164025 M^{6}}{1024 \ell^{4}} + \mathcal{O}(\ell_*^{-5}).
\end{align}
respectively.

Finally, we find the exact expression for the $q$-version of the shadow radius \cite{Perlick:2021aok} as
\begin{equation} \label{Rsh_ex}
    R_{\rm sh} =  \frac{b_{\rm crit} r_{\rm obs}}{q(r_{\rm obs})} \sqrt{A(q(r_{\rm obs}))}.
\end{equation}
Again, for $\ell_*/M \to 0$ and $\ell_*/M \to \infty$, we find
\begin{align} \label{Rsh_far}
    R_{\rm sh}^{\rm far} &\sim 3 \sqrt{3}\, M - 6 \sqrt{3\ell_*}\, M \left(\frac{1}{r_{\rm obs}}\right)^{1/2} + \mathcal{O}(r_{\rm obs}^{-3/2}), \nonumber \\
    R_{\rm sh}^{\rm far} &\sim \frac{3 \sqrt{3}\, M \,r_{\rm obs}^{3/2}}{8 \ell^{\frac{3}{2}}}-\frac{3 \sqrt{3}\, M r_{\rm obs}}{4 \ell} + \mathcal{O}(r_{\rm obs}^{1/2}).
\end{align}
which is the one we need to constraint $\ell_*$ using the EHT data/results for Sgr. A* and M87*. Indeed, due to the uncertainty of the photonsphere's position caused by the intrinsic stochasticity of spacetime, the behavior of the shadow radius is also affected. However, it is interesting to note how the multi-length scale is coupled to the observer's position relative to the black hole.

The Schwarzschild shadow radius is bounded by the following uncertainties: $4.209M \leq R_{\rm Sch} \leq 5.560M$ for Sgr. A$^*$ at $2\sigma$ level of significance \cite{Vagnozzi:2022moj}, and $ 4.313M \leq R_{\rm Sch} \leq 6.079M$ for M87$^*$ at $1\sigma$ level of significance \cite{EventHorizonTelescope:2021dqv}. Let $\delta$ represent these upper and lower bounds. Then, we have
\begin{equation}
    {}^q R_{\rm sh}^{\rm far} = R_{\rm Schw} + \delta.
\end{equation}
Here, we can find the parameter estimation for the multi-scale length $\ell_*$, again, for the cases $\ell_*/M \to 0$ and $\ell_*/M \to \infty$ as
\begin{align} \label{cons}
    \ell_* &\sim \frac{\delta^2 r_{\rm obs}^3 }{108M^4}, \nonumber \\
    \ell_* &\sim \frac{(3\sqrt{3}M)^{2/3} r_{\rm obs}}{4 \left(3\sqrt{3}M +\delta \right)^{2/3}},
\end{align}
respectively. For the second equation above, we only used the first term as this is the largest contribution.
It shows its dependence on the observer's distance from the black hole in terms of the usual $r$-coordinate and cannot be negative. Now, for Sgr. A* ($M = 6.40\times 10^{9} \text{ m}$), $\delta = \pm 0.987M$ and $r_{\rm obs} = 4.02 \times 10^{10} M$, giving $\ell_*/M = 5.86\times 10^{29}$. For M87* ($M = 9.60\times 10^{12} \text{ m}$), $\delta = \pm 0.883M$ and $r_{\rm obs} = 5.4 \times 10^{10} M$ , which leads to $\ell_*/M = 1.14\times 10^{30}$. 

For comparison, the cosmological horizon is only $\sim 9.51 \times 10^{25} \text{ m}$, and the obtained values of $\ell_*$ greatly exceed it. Furthermore, it turns out that the first expression in Eq. \eqref{cons} does not apply to astrophysical black holes since the condition $M \gg \ell_*$ is not met. However, if we examine the second expression, we find that $8.95 \times 10^{9} \leq \ell_*/M \leq 1.16 \times 10^{10}$ for Sgr. A*, and $1.22 \times 10^{10} \leq \ell_*/M \leq 1.53 \times 10^{10}$ for M87*. Here, we see from these results that the second expression in Eq. \eqref{cons} is the one applicable for constraining $\ell_*$ in astrophysical black holes since the condition that $\ell_* \gg M$ is met. The implication of this result is such that the shadow radius is completely modified as shown in the second expression in Eq. \eqref{Rsh_far}. Note that the Schwarzschild case is recovered if $\ell_* \sim 0.1164\, r_{\rm obs}$.

Finally, when the cosmological constant is viewed as an effect of a multi-scaling nature of geometry in the weighted derivative formalism, then Eq. \eqref{gamma} applies. Basically, the shadow analysis would be the same as that of the standard dS/AdS black hole spacetime, where the photonsphere is found not to be affected by the cosmological constant: $r_{\rm ph} = 3M$. Then, the expression for the critical impact parameter is
\begin{equation}
    b^2_{\rm crit} = \frac{54M^2}{1-9\chi M^2}.
\end{equation}
Again, as one follows the formalism of deriving the shadow radius, we find that in the far approximation is
\begin{equation}
    {}^w R_{\rm sh}^{\rm far} \sim 3 \sqrt{3}\, M -\frac{M \sqrt{3}\, \chi  \,r_{\rm obs}^{2}}{4} - \mathcal{O}(r_{\rm obs}).
\end{equation}
Then, using the EHT results, we can possibly constrain $\chi$:
\begin{equation}
    \chi = -\frac{4 \delta \sqrt{3}}{3 M r_{\rm obs}^2}.
\end{equation}
In dS spacetime where $\chi>0$, we choose $\delta < 0$, whereas, in AdS spacetime, $\delta > 0$ must be chosen. For Sgr. A*, we find that $\chi \sim 3.49\times 10^{-41} \text{ m}^{-2}$, while for M87*, $\chi \sim 7.59\times 10^{-48} \text{ m}^{-2}$. We see that constraints on M87* give a closer value to $\Lambda \sim 10^{-52} \text{ m}^{-2}$, hence, providing a small but still significant deviation. We may attribute the deviation to the multi-fractional effects of spacetime under the weighted derivative formalism.

\section{Deflection angle in the weak field regime} \label{sec3}
In this section, we will use the version of the Gauss-Bonnet theorem (GBT) that allows us to calculate the weak deflection angle even if the spacetime is non-asymptotically flat. To this aim, the $1+2$ dimensional metric in Eq. \eqref{q_met2} is utilized to the general version of GBT given by \cite{Li:2020wvn}:
\begin{equation} \label{wda_Li}
    \Theta = \iint_{_{q(r_{\rm ph})}^{\rm R }\square _{q(r_{\rm ph})}^{\rm S}}KdS + \phi_{\text{RS}}.
\end{equation}
Note that since we are dealing with multi-fractional theory, the GBT above must be rewritten in its $q$-version. The integral above is evaluated in a quadrilateral spanning from the position of the source S, the photonsphere $q(r_{\rm ph})$, the position of the receiver R, and back to the position of the source. Also, $\phi_{\rm RS}$ is the separation between the source and the receiver, as given by
\begin{equation}
    \phi_{\rm RS} = \phi_{\rm R} - \phi_{\rm S},
\end{equation}
where
$\phi_{\rm R} = \pi - \phi_{\rm S}$. Furthermore, $dS = \sqrt{g}dqd\phi$ is the infinitesimal curve surface and $g$ is the determinant of the Jacobi metric in its $q$-version as
\begin{align} \label{Jac_met}
    dl^2&=g_{ij}dx^{i}dx^{j}
    =(E^2-\mu^2A(q))\left[\frac{B(q)}{A(q)}dq^2+\frac{C(q)}{A(q)}d\phi^2\right], \nonumber \\
    g &= \frac{(E^2 - \mu^2 A(q))B(q)C(q)}{A(q)^2}.
\end{align}
In the above expression, $\mu$ is the mass of the time-like particle, and $E$ is the energy per unit mass. In terms of these variables, we can rewrite Eq. \eqref{wda_Li} as
\begin{equation} \label{wda_Li2}
    \Theta = \int^{\phi_{\rm R}}_{\phi_{\rm S}} \int_{q(r_{\rm ph})}^{q(r(\phi))} K\sqrt{g} \, dq \, d\phi + \phi_{\rm RS},
\end{equation}
where the Gaussian curvature is given by
\begin{equation} \label{G_curva}
    K=\frac{1}{\sqrt{g}}\left[\frac{\partial}{\partial\phi}\left(\frac{\sqrt{g}}{g_{qq}}\Gamma_{qq}^{\phi}\right)-\frac{\partial}{\partial q}\left(\frac{\sqrt{g}}{g_{qq}}\Gamma_{q\phi}^{\phi}\right)\right] 
    =-\frac{1}{\sqrt{g}}\left[\frac{\partial}{\partial q}\left(\frac{\sqrt{g}}{g_{qq}}\Gamma_{q\phi}^{\phi}\right)\right].
\end{equation}
The Jacobi metric implies that $\Gamma_{qq}^{\phi} = 0$, and using the photonsphere as the integration domain,
\begin{equation}
    \left[\int K\sqrt{g}dq\right]\bigg|_{q=q(r_{\rm ph})} = 0,
\end{equation}
resulting to
\begin{equation} \label{gct}
    \int_{q(r_{\rm ph})}^{q(r(\phi))} K\sqrt{g}dq = -\frac{A(q)\left(E^{2}-A(q)\right)C'-E^{2}C(q)A(q)'}{2A(q)\left(E^{2}-A(q)\right)\sqrt{B(q)C(q)}}\bigg|_{q = q(r(\phi))},
\end{equation}
where prime denotes differentiation with respect to $q$.

As indicated by $q(r(\phi))$, the orbit equation is needed. By allowing $u(q) = q^{-1}$,
\begin{equation}
    F(u) \equiv \left(\frac{du(q)}{d\varphi}\right)^2 
    = \frac{C(u(q))^2u(q)^4}{A(u(q))B(u(q))}\Bigg[\left(\frac{E}{J}\right)^2-A(u(q))\left(\frac{1}{J^2}+\frac{1}{C(u(q))}\right)\Bigg],
\end{equation}
where $J = Evb$ is the angular momentum of the particle, and $b$ is the impact parameter. The above results to
\begin{equation} \label{e_u(final)}
    {}^wu(\phi) = \frac{1}{b}\sin \! \left(\phi\right)+ \frac{M}{b^{2} v^{2}}\left(1 + v^2 \cos \! \left(\phi \right)\right).
\end{equation}
Using the above expression on Eq. \eqref{gct},
\begin{equation}
    \int_{q(r_{\rm ph})}^{q(r(\phi))} K\sqrt{g}dq \sim -1+\frac{\left(2 E^{2}-1\right) \sin \! \left(\phi \right) M}{b \left(E^{2}-1\right)}.
\end{equation}
Then,
\begin{equation}
    \int_{\phi_{\rm S}}^{\phi_{\rm R}} \int_{q(r_{\rm ph})}^{q(r(\phi))} K\sqrt{g} \, dq \, d\phi \sim -\phi_{\rm RS} -\frac{\left(2 E^{2}-1\right) M}{\left(E^{2}-1\right) b}\cos \! \left(\phi\right) \bigg\vert_{\phi_{\rm S}}^{\phi_{\rm R}}.
\end{equation}
We solve $\phi$ as,
\begin{equation}
    \phi = \arcsin \! \left(b u(q) \right)-\frac{M \left[v^{2} \left(b^{2} u(q)^{2}-1\right)-1\right]}{2 \sqrt{1-b^{2} u(q)^{2}}\, b \,v^{2}},
\end{equation}
and it follows that
\begin{equation}
    \cos(\phi) = \sqrt{1-b^{2} u(q)^{2}}+\frac{u(q) \left[v^{2} \left(b^{2} u(q)^{2}-1\right)-1\right] M}{2 \sqrt{1-b^{2} u(q)^{2}}\, v^{2}}.
\end{equation}
Due to the relations
\begin{equation}
    \cos \left(\pi - \phi\right) = -\cos \left(\phi\right), \qquad \phi_{\rm RS} = \pi - 2\phi,
\end{equation}
we can find the weak deflection angle, in its $q$-version as
\begin{equation}
    {}^q\Theta = \frac{2 \left(2 E^{2}-1\right) M \cos \! \left(\phi \right)}{b \left(E^{2}-1\right)} = \frac{2 \left(v^{2}+1\right) M \sqrt{1-b^{2} u(q)^{2}}}{b \,v^{2}}.
\end{equation}
Using Eq. \eqref{q}, we can finally recast the above equation as
\begin{equation}
    {}^q\Theta = \frac{2 \left(v^{2}+1\right) M}{b \,v^{2}} \sqrt{1-\frac{b^{2} u^{2}}{\left(2 \sqrt{lu}+1\right)^{2}}},
\end{equation}
where $u$ is now defined as $r^{-1}$. In the far approximation,
\begin{equation}
    {}^q\Theta^{\rm far}_{\rm timelike} \sim \frac{2 \left(v^{2}+1\right) M}{b \,v^{2}} \left(1-\frac{b^{2} u^{2}}{2}+2 b^{2} \sqrt{\ell_*}\, u^{\frac{5}{2}} -\mathcal{O}(u^{3}) \right).
\end{equation}
With photons, where $v = 1$,
\begin{equation} \label{wda_q}
    {}^q\Theta^{\rm far}_{\rm null} \sim \frac{4M}{b}\left( 1-\frac{b^{2} u^{2}}{2} + 2 b^{2} \sqrt{\ell_*}\, u^{\frac{5}{2}} -\mathcal{O}(u^{3}) \right),
\end{equation}
where $u$ can now be associated with $r_{\rm obs}$.

Let us constrain $\ell_*$ through Solar System test. If the photon just grazes the surface of the Sun, then $b = R_{\odot}$, and $M=M_{\odot}$. Results from the parametrized post-Newtonian (PPN) formalism equation for the deflection of light given as \cite{Chen:2023bao}
\begin{equation}
    \Theta^{\rm PPN} \backsimeq \frac{4M_{\odot}}{R_{\odot}}\left(\frac{1+\gamma}{2} \right).
\end{equation}
According to the astrometric observation of the Very Long Baseline Array (VLBA), $\gamma$ is the PPN deflection parameter, equal to $0.9998 \pm \Delta$, where $\Delta = 0.0003$. \cite{fomalont2009progress, Chen:2023bao}. Then, using Eq. \eqref{q} on Eq. \eqref{wda_q}, we find the relation
\begin{equation}
    \frac{4 M_{\odot}}{R}\left( 1-\frac{R^{2}}{2 r_{\rm obs}^{2}}+\frac{2 R^{2} \sqrt{l}}{r_{\rm obs}^{5/2}}
 \right) = \frac{4M_{\odot}}{R}\left(\frac{1.9998 \pm \Delta}{2} \right),
\end{equation}
Through comparison,
\begin{equation}
    \ell_* \sim \frac{\left( -0.0002 \pm \Delta \right)^{2} r_{\rm obs}^{5}}{16 R_{\odot}^{4}} + \mathcal{O}(r_{\rm obs}^3).
\end{equation}
Next, since $M_\odot = 1476.6148 \text{ m}$, $R_{\odot} = 696340000 \text{ m}$, and $r_{\rm obs} = 1.4899 \times 10^{11} \text{ m}$, we find constraints for $l_*$
\begin{equation}
    1.32 \times 10^{8} \leq \ell_*/M_{\odot} \leq 3.30 \times 10^{9}.
\end{equation}

In the weighted derivative formalism, where we used Eq. \eqref{gamma}, and find the weak deflection angle as
\begin{equation}
    {}^w\Theta = \frac{2 \left(v^{2}+1\right) M \sqrt{1-b^{2} u^{2}}}{v^{2} b}+\frac{b \left(v^{2}-2\right) \chi  \sqrt{1-b^{2} u^{2}}}{6 v^{2} u}.
\end{equation}
Now, if $u \to 0$,
\begin{equation}
    {}^w\Theta^{\rm far}_{\rm timelike} = \frac{2 \left(v^{2}+1\right) M}{v^{2} b}+\frac{b \left(v^{2}-2\right) \chi r_{\rm obs} }{6 v^{2}}.
\end{equation}
For photons,
\begin{equation}
    {}^w\Theta^{\rm far}_{\rm null} = \frac{4 M}{b}-\frac{b \chi  r_{\rm obs}}{6}
\end{equation}
Given the distance of Earth from the Sun, $r_{\rm obs} = 148.99$ million km, constraints for $\chi$ using Solar System test can be found as
\begin{equation}
    \chi = -\frac{12 M \left(- 0.0002 \pm \Delta \right)}{R_{\odot}^2 r_{\rm obs}},
\end{equation}
which yields the bounds
\begin{equation}
    -5.35 \times 10^{-23} \leq \chi M_{\odot}^2 \leq 2.67 \times 10^{-22}.
\end{equation}
The above are the only bounds to potentially detect the deviation caused by the weighted derivative formalism in the multi-fractional theory of spacetime. The weak gravitational field of the Sun, then, provides a significantly large deviation to the accepted value of the cosmological constant $\Lambda$. As a final remark, since the positive bound describes the dS spacetime, which is the one that is needed in reality.

\section{Conclusion} \label{conc}
In this work, how the multi-fractional theory, particularly through the formalism of q-derivatives and weighted derivatives is explored. By incorporating scale-dependent geometric structures, we found that the presence of a multi-scale length $\ell_*$ introduces notable modifications to the shadow radius and weak deflection angle. These deviations leave traces of the quantum gravitational effects in the context of multi-fractional theory in astrophysical settings.

The analysis of black hole shadows, particularly for the supermassive black holes M87* and Sgr A*, revealed constraints on $\ell_*$ that are of the order $10^{10}$, implying that multi-fractional effects, while difficult to detect, might be present at large scales. The weak deflection angle computed using the generalized Gauss-Bonnet theorem further supports the sensitivity of gravitational lensing to these multi-fractional corrections, offering a secondary observational test. It was shown that in the weak field regime, such as those in the Solar System, constraints on $\ell_*$ were similarly placed within tight bounds.

The findings of this work not only extend our understanding of black holes within the multi-fractional theory framework but also demonstrate that this approach can lead to testable predictions in gravitational lensing and black hole shadow phenomena. Future work should focus on refining observational techniques and applying these theoretical models to upcoming datasets from projects such as the Event Horizon Telescope. Additionally, the role of the cosmological constant and its interpretation within the weighted derivative formalism deserves further investigation, particularly in how it affects large-scale structures and horizon formation. The multi-fractional theory of spacetime then offers a promising pathway for bridging classical general relativity with quantum gravity, providing testable predictions for future astrophysical observations. While current constraints suggest that these effects are small, ongoing advancements in observational precision may soon reveal their measurable imprints on black hole environments.

In future works, it would be interesting to explore further the associated black hole geometry and properties where the most general $q$-version of the black hole spacetime is affected by a modulation term \cite{Calcagni:2017ymp}. Given those complications, the analysis is expected to be interestingly requires some numerical approach. While the $q$ derivatives are interesting in a phenomenological sense, $q$-versions of more general spacetime metrics can be studied, such as those of charged black holes, rotating black holes, etc.

\section{acknowledgements}
R. P. would like to acknowledge networking support of the COST Action CA18108 - Quantum gravity phenomenology in the multi-messenger approach (QG-MM), COST Action CA21106 - COSMIC WISPers in the Dark Universe: Theory, astrophysics and experiments (CosmicWISPers), the COST Action CA22113 - Fundamental challenges in theoretical physics (THEORY-CHALLENGES), and the COST Action CA21136 - Addressing observational tensions in cosmology with systematics and fundamental physics (CosmoVerse).

\bibliography{biblio}
\end{document}